\documentclass[onecollarge]{svjour2}       
\usepackage[russian,english]{babel}
\usepackage{graphicx}
\usepackage{amsmath}
\usepackage{cite}
\journalname{}
\begin{document}

\title{Revising the applicability of the Stone-von Neumann theorem to scattering a quantum particle on a one-dimensional potential barrier}


\author{N. L. Chuprikov 
}


\institute{N. L. Chuprikov \at
              Tomsk State Pedagogical University, 634041, Tomsk, Russia \\
              \email{chnl@tspu.edu.ru}           
}

\date{Received: date / Accepted: date}

\maketitle

\begin{abstract}

It is shown that the Stone-von Neumann theorem is inapplicable to scattering a quantum nonrelativistic particle on a one-dimensional
"short-range"\/ potential barrier, since the unboundedness of the position operator plays here a crucial role. The Shcr\"{o}dinger representation
associated with this process is reducible: long before and long after the scattering event the space of its asymptotes represents the direct sum
of the subspaces of left and right asymptotes. There is a dichotomous-context-induced superselection rule (SSR), in which the role of a
superselection operator is played by the Pauli matrix $\sigma_3$ and the role of superselection (coherent) sectors is played by the above
subspaces. By the SSR any superposition of states from different coherent sectors is a mixed state, and splitting the incident wave packet into
the transmitted and reflected parts is nothing but a conversion of a pure state into a mixed one. The average values of any observable can be
defined only for the transmission and reflection subprocesses. The former evolves within a single coherent sector of the Hilbert space, in the
momentum representation; while the latter evolves within a single coherent sector of this space, in the coordinate representation.

\keywords{one-dimensional scattering \and Stone-von Neumann theorem \and superselection rule}
\end{abstract}

\newcommand{\ppp}{\mbox{\hspace{5mm}}}
\newcommand{\ooo}{\mbox{\hspace{3mm}}}
\newcommand{\ooa}{\mbox{\hspace{1mm}}}
\newcommand{\ppd}{\mbox{\hspace{18mm}}}
\newcommand{\ppt}{\mbox{\hspace{34mm}}}
\newcommand{\ppo}{\mbox{\hspace{10mm}}}

\section{Introduction}

The main goal of this paper is revising the contemporary quantum model (CQM) of scattering a nonrelativistic particle on a one-dimensional (1D)
potential barrier with support within some bounded spatial interval. This model is presented in many textbooks on quantum mechanics as an example
of an internally consistent quantum theory. However, in fact, this is not the case.

Let $\hat{H}$ be Hamiltonian that describes scattering a particle on the "short-range"\/ potential $V(x)$ which is nonzero in the interval
$[-a,a]$ (in order to focus all our attention on the main issue, we will assume that $\hat{H}$ has no bound states), and $|\psi_0\rangle$ be the
initial state of a particle. Then, according to the CQM (see, e.g., \cite{Pri,Bri,Ree,Tay}), the basic properties of this one-particle process can
be expressed in the following statements:
\begin{itemize}
\item [a)] There are in and out asymptotes $|\psi_{in}\rangle$ and $|\psi_{out}\rangle$ such that the scattering state
$e^{-i\hat{H}t/\hbar}|\psi_0\rangle$ "interpolates"\/ between them. The left and right parts of each asymptote are localized in the
non-overlapping spatial regions that lie on different sides of the barrier -- long before and long after the scattering event, i.e., when
$t\to\mp\infty$, a particle does not interact with the "short-range"\/ potential.
\item [b)] Both asymptotes are uniquely determined by the initial state of a particle: $|\psi_{in}\rangle=\hat{\Omega}_+ |\psi_0\rangle$ and
$|\psi_{out}\rangle=\hat{\Omega}_- |\psi_0\rangle$, where $\hat{\Omega}_{\mp}=\lim_{t\to \pm\infty} e^{i\hat{H}t/\hbar}e^{-i\hat{H}_0t/\hbar}$ are
the in and out M{\o}ller wave operators; $\hat{H}_0$ is the free one-particle Hamiltonian.
\item [c)] The space ${\cal{H}}_{in}$ of in asymptotes and the space ${\cal{H}}_{out}$ of out asymptotes
coincide with each other: ${\cal{H}}_{in}={\cal{H}}_{out}={\cal{H}}_{as}$ (weak asymptotic completeness). Moreover, in the
case considered, that is, when $\hat{H}$ has no bound states, ${\cal{H}}_{as}$ spans the entire Hilbert space ${\cal{H}}$,
that is, ${\cal{H}}_{as}={\cal{H}}$.
\item [d)] The Shcr\"{o}dinger representation is irreducible and, thus, ${\cal{H}}_{as}$ can not be presented as a direct sum of
nontrivial subspaces which would be invariant for the position and momentum operators. This also means that any asymptotic state from
${\cal{H}}_{as}$ is a pure state, and a superposition of any two asymptotic states from ${\cal{H}}_{as}$ is another pure state from
${\cal{H}}_{as}$.
\item [e)] There is a linear unitary transformation $\textbf{S}=\hat{\Omega}_{-}^\dag\hat{\Omega}_{+}$ which
"correlates the past and future asymptotics of interacting histories" \cite{Ree}:
$|\psi_{out}\rangle=\textbf{S}|\psi_{in}\rangle$. "The fact that $\textbf{S}$ is unitary means that for every normalized
$|\psi_{in}\rangle$ there is a unique normalized $|\psi_{out}\rangle$ and vise versa; and also (because $\textbf{S}$ is
linear) that the correspondence between $|\psi_{in}\rangle$ and $|\psi_{out}\rangle$ preserves superposition, that is, if
$|\psi_{in}\rangle=a|\phi_{in}\rangle+b|\chi_{in}\rangle$, then
$|\psi_{out}\rangle=a|\phi_{out}\rangle+b|\chi_{out}\rangle$" \cite{Tay} p.36.
\end{itemize}

Note that the statement (d) is, perhaps, most fundamental here, because the irreducibility of the Shcr\"{o}dinger representation guarantees the
validity of the superposition principle in the space ${\cal{H}}$. Nevertheless, namely this statement makes this model internally inconsistent and
must be discarded from this list, because it contradicts the item (a) and is based on the erroneous assumption that the Stone-von Neumann theorem
is applicable to this scattering process.

As is known (see, e.g., \cite{Pri,Bri}), in the Stone-von Neumann theorem the position operator is treated like bounded operators. At the same
time the unboundedness of this operator plays a crucial role in the problem under study: the item (a) implies that, in the limits $t\to\pm\infty$,
a particle does not interact with the barrier and the space ${\cal{H}}_{as}$ represents a direct sum of the non-overlapping subspaces of left and
right asymptotes. Making use of Hall's terminology (see \cite{Bri} p.281), we can say that in the case of this process we meet just with that
"bad" case of the canonical commutation relations, which is not covered by the Stone-von Neumann theorem.

As will be shown below, the dichotomous physical context, formed by the different physical conditions on either side of the barrier, induces a
superselection rule (SSR), according to which the left and right asymptotes belong to different coherent (superselection) sectors of the Hilbert
space ${\cal{H}}_{as}$ when $t\to\pm\infty$. These subspaces are invariant for the position and momentum operators, as well as for the Pauli
matrix $\sigma_3$ that plays in this space the role of a superselection operator. By this rule a superposition of left and right asymptotes
represents a mixed state, rather than a pure one.




\section{Stationary states of a particle in the formalism of the transfer and scattering matrices.} \label{station}

We begin our analysis with solving the stationary Schr\"{o}dinger equation for a particle with a given energy $E=(\hbar k)^2/2m$; $m$ is its mass.
In the general case the wave function $\Psi(x;k)$, beyond the interval $[-a,a]$, can be written in the form
\begin{eqnarray} \label{1}
\Psi(x;k)=\left\{
\begin{array}{cc}
A_{L,in}(k)\ooa e^{ikx}+A_{L,out}(k)\ooa e^{-ikx},\ppp x\leq -a; \\
A_{R,out}(k)\ooa e^{ikx}+A_{R,in}(k)\ooa e^{-ikx},\ppp x\geq +a
\end{array}\right.
\end{eqnarray}
Its amplitudes in the regions $x\leq -a$ and $x\geq a$ are linked by the transfer matrix $\textbf{Y}(k)$:
\begin{eqnarray} \label{111}
\left(\begin{array}{cc} A_{L,in} \\ A_{L,out} \end{array} \right)=\textbf{Y} \left(\begin{array}{cc} A_{R,out} \\ A_{R,in}
\end{array} \right);\ppp \textbf{Y}=\left(
\begin{array}{cc}
q & p \\
p^* & q^*
\end{array} \right);
\end{eqnarray}
where $q(-k)=q^*(k)$, $p(-k)=p^*(k)$. According to \cite{Ch8}, for any potential barrier with support inside $[x_1,x_2]$ the transfer-matrix
elements can be written as follows,
\begin{eqnarray} \label{2}
q=\frac{1}{\sqrt{T(k)}}\ooa e^{i\left[k(x_2-x_1)-J(k)\right]},\ppp p=i\sqrt{\frac{R(k)}{T(k)}}\ooa
e^{i[-k(x_2+x_1)+F(k)]},\ppp R=1-T;
\end{eqnarray}
$T(-k)=T(k)$, $J(-k)=-J(k)$, $F(-k)=\pi-F(k)$; for the case considered $x_2-x_1=d=2a$ and $x_2+x_1=0$. For any symmetric
potential barrier, $V(-x)=V(x)$, the phase $F$ takes only two values: $0$ or $\pi$. In this case, a piecewise-constant
function $F(k)$ has discontinuities at the points where the reflection coefficient equals to zero.

Note that the scattering parameters (the transmission $T$ and reflection $R$ coefficients, as well as the phases $J$ and
$F$) can be calculated (analytically or numerically) for potential barriers of any form. For this purpose one can use
either analytical expressions in \cite{Ch8}, if $V(x)$ is the rectangular potential barrier or the $\delta$-potential, or
recurrence relations, if $V(x)$ represents a system of $\delta$-potentials and piecewise continuous potential barriers.
Thus, we can further assume that the matrix $\textbf{Y}(k)$ and the scattering matrix $\textbf{S}$ that links the
amplitudes $A_{L,out}$ and $A_{R,out}$ of outgoing waves with the amplitudes $A_{L,in}$ and $A_{R,in}$ of incoming waves
are known.

Since this link can be realized in two ways, and both will be important for our approach (see Section \ref{colomn}), we
consider two scattering matrices -- $\textbf{S}_k$ and $\textbf{S}_x$:
\begin{eqnarray} \label{3}
\left(\begin{array}{cc} A_{R,out} \\ A_{L,out} \end{array} \right)=\textbf{S}_k \left(\begin{array}{cc} A_{L,in} \\
A_{R,in}
\end{array} \right),\ooa \textbf{S}_k=\frac{1}{q}\left(
\begin{array}{cc}
1 & -p \\
p^* & 1
\end{array} \right);\ooo
\left(\begin{array}{cc} A_{L,out} \\ A_{R,out} \end{array} \right)=\textbf{S}_x \left(\begin{array}{cc} A_{L,in} \\
A_{R,in}
\end{array} \right),\ooa \textbf{S}_x=\frac{1}{q}\left(
\begin{array}{cc}
p^* & 1 \\
1 & -p
\end{array} \right)
\end{eqnarray}
It is assumed that the amplitudes $A_{L,in}(k)$ and $A_{R,in}(k)$ are independent and equal to zero for $k\leq 0$ as well as obey the condition
$|A_{L,in}(k)|^2+|A_{R,in}(k)|^2=1$. When changing the sign of the wave number $k$, the incoming and outgoing waves change roles:
$A_{L,in}(-k)\equiv A_{L,out}^\prime(k)$, $A_{R,in}(-k)\equiv A_{R,out}^\prime(k)$, $A_{L,out}(-k)\equiv A_{L,in}^\prime(k)$, $A_{R,out}(-k)\equiv
A_{R,in}^\prime(k)$. The primed variables are linked by the relations (\ref{3}).

Our next step is to determine the spaces of in and out asymptotes -- freely moving wave packets that describe non-stationary localized (physical)
states of a particle in the limits $t\to -\infty$ and $t\to +\infty$, respectively (that is, at the initial and final stages of scattering).

\section{The space of in and out asymptotes in the formalism of the scattering matrix}

Note, in the general case the in asymptote has two components. That is, it represents a superposition of the left and
right asymptotes (wave packets) moving toward the barrier in the region $(-\infty,-a)$ and $(a,\infty)$, respectively. The
same concerns the out asymptote, but its left and right asymptotes move away from the barrier.

Let $A_{L,out}$ and $A_{R,in}$ be the amplitudes of waves that move, on the $OX$-axis, from the right to the left. Then
$$\Psi_{L,in}(k,t)=A_{L,in}(k)e^{-iE(k)t/\hbar},\ooo \Psi_{R,in}(k,t)=A_{R,in}(-k)e^{-iE(k)t/\hbar}$$ be wave packets
moving toward the barrier in the remote regions $\cal{A}$ and $\cal{B}$ that lie in the intervals $(-\infty,-a)$ and $(a,\infty)$, respectively;
and
$$\Psi_{L,out}(k,t)=A_{L,out}(-k)e^{-iE(k)t/\hbar},\ooo \Psi_{R,out}(k,t)=A_{R,out}(k)e^{-iE(k)t/\hbar}$$ be wave packets
moving away from the barrier in the regions $\cal{A}$ and $\cal{B}$, respectively.

According to the item (a), in the limits $t\to\pm\infty$, the left and right components of in and out asymptotes are
nonzero, in the $x$-space, within the nonoverlapping intervals $(-\infty,-a)$ and $(a,\infty)$. This means that in these
limiting cases the space ${\cal{H}}_{as}$ represents a direct sum of the subspaces ${\cal{H}}_{L}$ and ${\cal{H}}_{R}$ of
left and right asymptotes, respectively: ${\cal{H}}={\cal{H}}_{as}={\cal{H}}_{L}\oplus{\cal{H}}_{R}$. And, if this is so
in the $x$-representation, this must be also valid in the $k$-representation. The item (a) also implies that the left and
right asymptotes are well localized, both in the $x$- and $k$-spaces: they are such that the average values of the
operators $\hat{X}^n$ and $\hat{P}^n$ exist for any value of $n$; here $\hat{X}$ and $\hat{P}$ are the position and
momentum operators, respectively.


To construct asymptotes with such properties in the $k$-space, we will assume that the functions $A_{L,in}(k)$ and
$A_{R,in}(k)$ belong to the spaces ${\cal{S}}(\Omega_k^+)$ and ${\cal{S}}(\Omega_k^-)$, respectively; where
${\cal{S}}(\Omega_k^+)$ is the Schwartz subspace of infinitely differentiable functions which are zero on the semiaxis
$(-\infty,0]$ and diminish in the limit $k\to\infty$ more rapidly than any power function; ${\cal{S}}(\Omega_k^-)$ is the
Schwartz subspace of infinitely differentiable functions which are zero on the semiaxis $[0,+\infty)$ and diminish in the
limit $k\to-\infty$ more rapidly than any power function. Thus, the wave packets $\Psi_{L,in}(k,t)$ and
$\Psi_{R,out}(k,t)$ belong to the subspace ${\cal{S}}(\Omega_k^+)$, while $\Psi_{L,out}(k,t)$ and $\Psi_{R,in}(k,t)$
belong to the subspace ${\cal{S}}(\Omega_k^-)$. And, as a consequence, in the $k$-representation, the space ${\cal{H}}$,
in the limits $t\to\pm\infty$, represents the sum of the non-overlapping subspaces ${\cal{S}}(\Omega_k^-)$ and
${\cal{S}}(\Omega_k^+)$: ${\cal{H}}={\cal{H}}_{as}={\cal{S}}(\Omega_k^-)\oplus{\cal{S}}(\Omega_k^+)$.

However, we have also to take into account that the left and right asymptotes from these two $k$-subspaces must be such
that their Fourier images are localized, in the $x$-space, in the remote non-overlapping spatial regions $\cal{A}$ and
$\cal{B}$. In order to ensure this property for both stages of scattering, it is sufficient to construct the space of in
asymptotes with such a property (the corresponding out asymptotes can be found with making use of the scattering matrix).

Let $A_{(+)}(k)$ and $A_{(-)}(k)$ be such wave packets from ${\cal{S}}(\Omega_k^+)$ and ${\cal{S}}(\Omega_k^-)$, respectively, that their "centers
of mass" (CMs) are positioned at the origin of coordinates. Then the needed left and right in asymptotes, in the $k$-representation, can be
written as $A_{L,in}(k)=A_{(+)}(k)e^{ikD}$ and $A_{R,in}(k)=A_{(-)}(k)e^{-ikD}$, respectively; here $D$ is the distance between the CM of each
wave packet and the origin of coordinates. For example, we can take as the left and right in asymptotes the following two wave functions:
$A_{L,in}(k)=N\cdot \exp\left[-\frac{l(k-k_0)^2}{k}+ikD\right]$ for $k\in (0,\infty)$ and $A_{R,in}(k)=N\cdot
\exp\left[\frac{l(k+k_0)^2}{k}-ikD\right]$ for $k\in (-\infty,0)$; here $N$ is the normalization constant, $k_0$ and $l$ are positive parameters.
To ensure the localization of the left and right in asymptotes in the remote spatial regions $\cal{A}$ and $\cal{B}$, respectively, we have to
consider these expressions in the limit $D\to\infty$.

So, the in asymptotes $\Psi_{L,in}$ and $\Psi_{R,in}$, as well as the corresponding out asymptotes $\Psi_{L,out}$ and $\Psi_{R,out}$ do not
overlap each other both in the $k$-space and in the $x$-space:
\begin{eqnarray*}
\Psi_{L,in}(k,t),\ooa \Psi_{R,out}(k.t)\in {\cal{S}}(\Omega_k^+),\ooo \Psi_{R,in}(k,t),\ooa\Psi_{L,out}(k,t)\in {\cal{S}}(\Omega_k^-),\\
\Psi_{R,in}(x,t),\ooa\Psi_{R,out}(x,t)\in {\cal{S}}(\Omega_x^+),\ooo \Psi_{L,in}(x,t),\ooa \Psi_{L,out}(x.t)\in {\cal{S}}(\Omega_x^-).
\end{eqnarray*}

\section{On the rigged (equipped) Hilbert space associated with the process} \label{rigg}

Note that in a more accurate description of state spaces in quantum mechanics (see, e.g., \cite{Rob,Mad,Madr}) the states of a particle involved
in this scattering process form the rigged (equipped) Hilbert space ${\cal{H}}^{rig}$ -- a Gelfand triplet $\Phi \subset {\cal{H}} \subset
\Phi^\times$, where ${\cal{H}}$ is a Hilbert space; $\Phi$ is the space of 'physical states'; $\Phi^\times $ is the space of antilinear
functionals over $\Phi$, which includes right eigenvectors of one-particle operators $\hat{X}$ and $\hat{P}$ (the corresponding {\it bra}-vectors
belong to the space $\Phi^\prime$ of linear functionals over $\Phi$). The term 'physical states' implies that for such states expectation values
exist for any finite degree of the operators $\hat{X}$ and $\hat{P}$. According to \cite{Rob,Mad,Madr}, such states belong to the Schwartz space
${\cal{S}}$ which is invariant with respect to the Fourier-transform.

The space of asymptotes should be denoted now as $\Phi_{as}$, and namely $\Phi_{as}$ spans the whole space $\Phi$ when there are no bound states.
That is, now the space $\Phi$ has (together with $\Phi_{as}$) a nontrivial structure in the limits $t\to\pm\infty$. If one considers asymptotes in
the $k$-representation, $\Phi_{as}=\Phi_{as}(\Omega_k^-)\oplus\Phi_{as}(\Omega_k^+)$ where $\Phi_{as}(\Omega_k^-)={\cal{S}}(\Omega_k^-)$ and
$\Phi_{as}(\Omega_k^+)={\cal{S}}(\Omega_k^+)$. While in the $x$-representation $\Phi_{as}=\Phi_{as}(\Omega_x^-)\oplus\Phi_{as}(\Omega_x^+)$ where
$\Phi_{as}(\Omega_x^-)={\cal{S}}(\Omega_x^-)$ and $\Phi_{as}(\Omega_x^+)={\cal{S}}(\Omega_x^+)$. Thus, there are reasons to believe that the space
${\cal{H}}^{rig}$ has, too, a more complex structure than it was assumed in \cite{Rob,Mad,Madr}.

\section{Asymptotic states as two-component wave functions} \label{colomn}

Since the left and right components of the in asymptote $|\Psi_{in}\rangle=|\Psi_{L,in}\rangle+|\Psi_{R,in}\rangle$ (in the limit $t\to -\infty$)
and out asymptote $|\Psi_{out}\rangle=|\Psi_{L,out}\rangle+|\Psi_{R,out}\rangle$ (in the limit $t\to +\infty$) belong to the non-overlapping
spaces, their scalar products equal to zero and the expressions for the norms of the vectors $|\Psi_{in}\rangle$ and $|\Psi_{out}\rangle$ do not
contain interference terms. That is,
$\langle\Psi_{in}|\Psi_{in}\rangle=\langle\Psi_{L,in}|\Psi_{L,in}\rangle+\langle\Psi_{R,in}|\Psi_{R,in}\rangle=1$ and
$\langle\Psi_{out}|\Psi_{out}\rangle=\langle\Psi_{L,out}|\Psi_{L,out}\rangle+\langle\Psi_{R,out}|\Psi_{R,out}\rangle=1$.

The scattering matrix formalism prompts us that the two-component in and out asymptotes can be presented, similarly to the Pauli spinor, in the
form of two-component columns. Thus, we will believe further that any two asymptotes $|\chi\rangle$ and $|\psi\rangle$ can be written as
$\left(\begin{array}{cc} \chi_1
\\ \chi_2
\end{array} \right)$ and $\left(\begin{array}{cc} \psi_1
\\ \psi_2 \end{array} \right)$, respectively, and their norms and scalar products are defined as
$\langle\chi|\chi\rangle=\langle\chi_1|\chi_1\rangle+\langle\chi_2|\chi_2\rangle$,
$\langle\psi|\psi\rangle=\langle\psi_1|\psi_1\rangle+\langle\psi_2|\psi_2\rangle$,
$\langle\chi|\psi\rangle=\langle\chi_1|\psi_1\rangle+\langle\chi_2|\psi_2\rangle$.

Note, the conformity between the components of any asymptote and the corresponding column depends on the scattering matrix taken as the basis for
the transition to the "two-column representation". In the $k$-space we have to use the formalism of the scattering matrix $\textbf{S}_k$, while in
the $x$-space we have to use the formalism of the scattering matrix $\textbf{S}_x$. And of importance is to stress that we can use {\it either}
the $k$-representation {\it or} the $x$-representation!

\subsection{$k$-representation}

Note that the matrix $\textbf{S}_k$ acts in the space of columns whose first elements describe waves moving along the $OX$-axis from the left to
the right, while the second elements describe waves moving in the opposite direction. In other words, the first elements of such columns are wave
functions that belong to the subspace ${\cal{S}}(\Omega_k^+)$, while the second ones are wave functions from ${\cal{S}}(\Omega_k^-)$. Thus, these
asymptotes can be presented in the form
$|\Psi_{in}\rangle=\left(\begin{array}{cc} \Psi_{L,in} \\
\Psi_{R,in}
\end{array} \right)$, $|\Psi_{out}\rangle=\left(\begin{array}{cc} \Psi_{R,out}
\\ \Psi_{L,out} \end{array} \right)$. The corresponding bra-vectors represent rows: $\langle\Psi_{in}|=\left(\Psi^*_{L,in}, \Psi^*_{R,in}\right)$,
$\langle\Psi_{out}|=\left(\Psi^*_{R,out}, \Psi^*_{L,out} \right)$.

Let us consider such a pair of vectors $|\phi_{k'}^{(1)}\rangle$ and $|\phi_{k'}^{(2)}\rangle$ with the parameter $k'>0$, as well as a pair of
vectors $|\phi_{x'}^{(1)}\rangle$ and $|\phi_{x'}^{(2)}\rangle$ with a given parameter $x'$ ($|x'|>a$) , that
\begin{eqnarray*}
\phi_{k'}^{(1)}(k)=\left(\begin{array}{cc} \delta(k-k')
\\ 0 \end{array} \right),\ooo \phi_{k'}^{(2)}(k)=\left(\begin{array}{cc} 0 \\ \delta(k+k')
\end{array} \right);\ooo \phi_{x'}^{(1)}(k)=\left(\begin{array}{cc} e^{ikx'}
\\ 0 \end{array} \right),\ooo \phi_{x'}^{(2)}(k)=\left(\begin{array}{cc} 0 \\ e^{-ikx'} \end{array} \right).
\end{eqnarray*}
The first pair among them gives the eigenvectors of the operator $\hat{P}$; with $\hat{P}|\phi_{k'}^{(1)}\rangle=+\hbar
k'|\phi_{k'}^{(1)}\rangle$, $\hat{P}|\phi_{k'}^{(2)}\rangle=-\hbar k'|\phi_{k'}^{(2)}\rangle$, and
$\langle\phi_{k'}^{(1)}|\phi_{k'}^{(2)}\rangle=0$. While the second pair gives the eigenvectors of the position operator $\hat{X}=i\frac{d}{dk}$;
with $\hat{X}|\phi_{x'}^{(1)}\rangle=-x'|\phi_{x'}^{(1)}\rangle$, $\hat{X}|\phi_{x'}^{(2)}\rangle=+x'|\phi_{x'}^{(2)}\rangle$, and
$\langle\phi_{x'}^{(1)}|\phi_{x'}^{(2)}\rangle=0$.

Note that, in the $k$-representation, the eigenvectors $\phi_{k'}^{(1)}(k)$ and $\phi_{k'}^{(2)}(k)$ of the momentum operator, corresponding to
its different eigenvalues, belong also to the different sectors ${\cal{S}}(\Omega_k^+)$ and ${\cal{S}}(\Omega_k^-)$. While, for example, the
eigenvectors $|\phi_{x'}^{(1)}\rangle$ and $|\phi_{-x'}^{(1)}\rangle$, corresponding to the different eigenvalues of the position operator, belong
to the same sector ${\cal{S}}(\Omega_k^+)$.

The stationary in and out asymptotes with a given positive value $k'$ can be written now in the form
$\Psi_{in}(k;k')=A_{L,in}(k)\phi_{k'}^{(1)}(k)+A_{R,in}(-k)\phi_{k'}^{(2)}(k)$ and
$\Psi_{out}(k;k')=A_{R,out}(k)\phi_{k'}^{(1)}(k)+A_{L,out}(-k)\phi_{k'}^{(2)}(k)$.

\subsection{$x$-representation}

In the $x$-representation we use the matrix $\textbf{S}_x$, because now
$|\Psi_{in}\rangle=\left(\begin{array}{cc} \Psi_{L,in} \\
\Psi_{R,in}
\end{array} \right)$ and $|\Psi_{out}\rangle=\left(\begin{array}{cc} \Psi_{L,out}
\\ \Psi_{R,out} \end{array} \right)$. That is, $\textbf{S}_x$ acts in the space of columns whose first elements
describe waves that constitute wave packets moving in the spatial region $x<-a$, while their second elements describe waves that constitute wave
packets moving in the region $x>+a$. Now, the first elements of columns are functions from the subspace ${\cal{S}}(\Omega_x^-)$, while the second
elements are functions from the subspace ${\cal{S}}(\Omega_x^+)$.

Let us consider such a pair of vectors $|\chi_{k'}^{(1)}\rangle$ and $|\chi_{k'}^{(2)}\rangle$ with a given parameter $k'$, as well as such a pair
of vectors $|\chi_{x'}^{(1)}\rangle$ and $|\chi_{x'}^{(2)}\rangle$ with the parameter $x'>0$, that
\begin{eqnarray*}
\chi_{k'}^{(1)}(x)=\left(\begin{array}{cc} e^{ik'x}
\\ 0 \end{array} \right),\ooo \chi_{k'}^{(2)}(x)=\left(\begin{array}{cc} 0 \\ e^{-ik'x} \end{array} \right);\ooo
\chi_{x'}^{(1)}(x)=\left(\begin{array}{cc} \delta(x+x')
\\ 0 \end{array} \right),\ooo \chi_{x'}^{(2)}(x)=\left(\begin{array}{cc} 0 \\ \delta(x-x')
\end{array} \right).
\end{eqnarray*}
The first pair of vectors gives the eigenvectors of the operator $\hat{P}=-i\hbar \frac{d}{dx}$; with $\hat{P}|\chi_{k'}^{(1)}\rangle=+\hbar
k'|\chi_{k'}^{(1)}\rangle$, $\hat{P}|\chi_{k'}^{(2)}\rangle=-\hbar k'|\chi_{k'}^{(2)}\rangle$, and
$\langle\chi_{k'}^{(1)}|\chi_{k'}^{(2)}\rangle=0$. While the second pair gives the eigenvectors of the operator $\hat{X}=x$: we have
$\hat{X}|\chi_{x'}^{(1)}\rangle=-x'|\chi_{x'}^{(1)}\rangle$, $\hat{X}|\chi_{x'}^{(2)}\rangle=+x'|\chi_{x'}^{(2)}\rangle$, and
$\langle\chi_{x'}^{(1)}|\chi_{x'}^{(2)}\rangle=0$.

Similarly, in the $x$-representation, the eigenvectors $\chi_{x'}^{(1)}(x)$ and $\chi_{x'}^{(2)}(x)$ of the position operator, corresponding to
its different eigenvalues, belong to the different sectors ${\cal{S}}(\Omega_x^-)$ and ${\cal{S}}(\Omega_x^+)$. While, for example, the
eigenvectors $|\phi_{k'}^{(1)}\rangle$ and $|\phi_{-k'}^{(1)}\rangle$, corresponding to the different eigenvalues of the momentum operator, belong
to the same sector ${\cal{S}}(\Omega_x^-)$.

The stationary in and out states for a given positive value $k$ can now be written in the form
$\Psi_{in}(x;k)=A_{L,in}(k)\chi_{k}^{(1)}(x)+A_{R,in}(k)\chi_{k}^{(2)}(x)$ and
$\Psi_{out}(x;k)=A_{L,out}(k)\chi_{-k}^{(1)}(x)+A_{R,out}(k)\chi_{-k}^{(2)}(x)$.

\section{The Pauli matrix $\sigma_3$ as a superselection operator in the space of asymptotes}

\subsection{$k$-representation}

Note that the vectors $|\phi_{k'}^{(1)}\rangle$, $|\phi_{k'}^{(2)}\rangle$, $|\phi_{x'}^{(1)}\rangle$ and $|\phi_{x'}^{(2)}\rangle$ are also the
eigenvectors of the Pauli matrix $\sigma_3=\left(\begin{array}{cc} 1 & 0 \\ 0 & -1
\end{array} \right)$. Indeed,
\begin{eqnarray*}
\sigma_3|\phi_{k'}^{(1)}\rangle=|\phi_{k'}^{(1)}\rangle,\ooo \sigma_3|\phi_{k'}^{(2)}\rangle=-|\phi_{k'}^{(2)}\rangle;\ppp
\sigma_3|\phi_{x'}^{(1)}\rangle=|\phi_{x'}^{(1)}\rangle,\ooo \sigma_3|\phi_{x'}^{(2)}\rangle=-|\phi_{x'}^{(2)}\rangle.
\end{eqnarray*}
Thus, though $\hat{X}$ and $\hat{P}$ do not commute with each other, each of them commutes with the operator $\sigma_3$. It is also important to
stress (see \cite{Hor2}) that the operator $\sigma_3$ can be expressed via the projection operators
$P_+= \left(\begin{array}{cc} 1 & 0 \\
0 & 0 \end{array} \right)$ è $P_-= \left(\begin{array}{cc} 0 & 0 \\ 0 & 1 \end{array} \right)$. Namely, $\sigma_3=P_+-P_-$.

Since the state space of a particle involved in the 1D scattering process is complete (see, e.g., \cite{Madr}) (and no matter which stage of this
process is regarded), the self-adjoint operator $\sigma_3$ can be treated (see \cite{Hor2,Hor3}) as the superselection operator which divides the
state space ${\cal{H}}^{rig}_{as}$, in the $k$-representation, into two coherent sectors (${\cal{H}}^{rig}_{as}$ is the rigged Hilbert space
${\cal{H}}^{rig}$ in the limits $t\to\mp\infty$):
\begin{eqnarray} \label{12}
{\cal{H}}^{rig}_{as}={\cal{H}}^{rig}_{as}(\Omega_k^+)\oplus {\cal{H}}^{rig}_{as}(\Omega_k^-);
\end{eqnarray}
${\cal{H}}^{rig}_{as}(\Omega_k^+)=\Phi_{as}(\Omega_k^+) \subset L_2^{as}(\Omega_k^+) \subset \Phi_{as}^\times(\Omega_k^+)$ and
${\cal{H}}^{rig}_{as}(\Omega_k^-)=\Phi_{as}(\Omega_k^-) \subset L_2^{as}(\Omega_k^-) \subset \Phi_{as}^\times(\Omega_k^-)$.

Here the subspace ${\cal{H}}_{as}^{rig}(\Omega_k^+)$ represents the coherent sector (let's call it the 'top coherent sector') that corresponds to
the eigenvalue $+1$ of the operator $\sigma_3$, while the subspace ${\cal{H}}_{as}^{rig}(\Omega_k^-)$ is the 'lower coherent sector' corresponding
to the eigenvalue $-1$. It is evident that $|\phi_{k'}^{(1)}\rangle,\ooa |\phi_{x'}^{(1)}\rangle\in \Phi_{as}^\times(\Omega_k^+)$, and
$|\phi_{k'}^{(2)}\rangle,\ooa |\phi_{x'}^{(2)}\rangle\in \Phi_{as}^\times(\Omega_k^-)$.

\subsection{$x$-representation}

Now the vectors $|\chi_{k'}^{(1)}\rangle$, $|\chi_{k'}^{(2)}\rangle$, $|\chi_{x'}^{(1)}\rangle$ and $|\chi_{x'}^{(2)}\rangle$ are eigenvectors of
the matrix $\sigma_3$:
\begin{eqnarray*}
\sigma_3|\chi_{k'}^{(1)}\rangle=|\chi_{k'}^{(1)}\rangle,\ooo \sigma_3|\chi_{k'}^{(2)}\rangle=-|\chi_{k'}^{(2)}\rangle;\ppp
\sigma_3|\chi_{x'}^{(1)}\rangle=|\chi_{x'}^{(1)}\rangle,\ooo \sigma_3|\chi_{x'}^{(2)}\rangle=-|\chi_{x'}^{(2)}\rangle.
\end{eqnarray*}
That is, as is expected, two coherent sector arise. In the $x$-representation,
\begin{eqnarray} \label{13}
{\cal{H}}^{rig}_{as}={\cal{H}}^{rig}_{as}(\Omega_x^-)\oplus {\cal{H}}^{rig}_{as}(\Omega_x^+);
\end{eqnarray}
\begin{eqnarray*}
{\cal{H}}^{rig}_{as}(\Omega_x^-)=\Phi_{as}(\Omega_x^-) \subset L_2^{as}(\Omega_x^-) \subset \Phi_{as}^\times(\Omega_x^-),\ooo
{\cal{H}}^{rig}_{as}(\Omega_x^+)=\Phi_{as}(\Omega_x^+) \subset L_2^{as}(\Omega_x^+) \subset \Phi_{as}^\times(\Omega_x^+).
\end{eqnarray*}
Now the eigenvalue $+1$ of the superselection operator $\sigma_3$ is associated with functions of the subspace ${\cal{H}}_{as}^{rig}(\Omega_x^-)$
(let's call it the 'left coherent sector', while the eigenvalue $-1$ is associated with functions that belong to the subspace
${\cal{H}}_{as}^{rig}(\Omega_x^+)$ (let's call it the 'right coherent sector'). It is evident that $|\chi_{k'}^{(1)}\rangle,\ooa
|\chi_{x'}^{(1)}\rangle\in \Phi_{as}^\times(\Omega_x^-)$ and $|\chi_{k'}^{(2)}\rangle,\ooa |\chi_{x'}^{(2)}\rangle\in
\Phi_{as}^\times(\Omega_x^+)$.

\section{Superselection rule and superposition principle for the scattering process} \label{pure}

According to the theory of SSRs \cite{Par1,Par2,Par3,Hor1,Hor2,Hor3}, any superposition of pure states from the same coherent sector gives another
pure state in this sector, while any superposition of pure states from different coherent sectors represents a mixed state. Thus, any
superposition of states from the coherent sectors $\Phi_{as}(\Omega_x^-)$ and $\Phi_{as}(\Omega_x^+)$, in the $x$-representation (or from the
coherent sectors $\Phi_{as}(\Omega_k^-)$ and $\Phi_{as}(\Omega_k^+)$, in the $k$-representation) represents a mixed state. Born's interpretation
of pure states is inapplicable to such a superposition -- its squared modulus cannot be treated as the probability distribution in the $x$-space
(or in the $k$-space), and the average value of any observable can not be defined for such a superposition.

One distinctive feature of such superpositions can be illustrated by the following example. Let $\hat{O}$ be a self-adjoint operator and
$|\Psi_{\cal{A}}\rangle$ and $|\Psi_{\cal{B}}\rangle$ be, respectively, the left and right in asymptotes which evolve in the remote spatial
regions ${\cal{A}}$ and ${\cal{B}}$; that is, in the $x$-space $|\Psi_{\cal{A}}\rangle\in \Phi_{as}(\Omega_x^-)$ and $|\Psi_{\cal{B}}\rangle\in
\Phi_{as}(\Omega_x^+)$. Besides, let $|\psi_\lambda\rangle =|\Psi_{\cal{A}}\rangle+e^{i\lambda}|\Psi_{\cal{B}}\rangle$, $|\psi_\nu\rangle
=|\Psi_{\cal{A}}\rangle+e^{i\nu}|\Psi_{\cal{B}}\rangle$; $\lambda$ and $\nu$ are real phases. Then
\begin{eqnarray*}
\langle\psi_\lambda|\hat{O}|\psi_\lambda\rangle=\langle\psi_\nu|\hat{O}|\psi_\nu\rangle=
\langle\Psi_{\cal{A}}|\hat{O}|\Psi_{\cal{A}}\rangle+\langle\Psi_{\cal{B}}|\hat{O}|\Psi_{\cal{B}}\rangle.
\end{eqnarray*}
Thus, at the initial stage of scattering the phases $\lambda$ and $\nu$ are unobservable quantities. This property is one of signs (see
\cite{Ear}) that the states $|\psi_\lambda\rangle$ and $|\psi_\nu\rangle$, representing the coherent superpositions of the left and right
asymptotes, are mixed states.

Of importance is also to stress that, according to probability theory, different physical conditions (contexts) in the remote spatial regions
${\cal{A}}$ and ${\cal{B}}$ "prepare"\/ {\it different} statistical ensembles of particles. As is stressed in \cite{Khr}, "Two collectives of
particles moving under two macroscopically distinct contexts form two different statistical ensembles"; and {\it "probabilistic data generated by
a few collectives\ldots cannot be described by a single Kolmogorov space" (ibid)} (see also \cite{Acc}).

Thus, according to the found SSR, any superposition of $|\Psi_{\cal{A}}\rangle$ and $|\Psi_{\cal{B}}\rangle$ can not be associated with a single
(pure) quantum ensemble. As a consequence, the so called bilateral scattering should be considered as a "mixture" of two subprocesses, each
representing the unilateral scattering.

\section{The unilateral scattering as a process of conversion of a pure state into a mixed one}

In the general case the scattering matrixes (\ref{3}) do not commute with the superselection operator $\sigma_3$:
\begin{eqnarray*}
[\textbf{S}_k,\sigma_3]=\frac{2}{q}\left(
\begin{array}{cc}
0 & p \\
p^* & 0
\end{array} \right)=2\sqrt{R(k)}e^{i(J(k)-kd)}\left(
\begin{array}{cc}
0 & e^{iF(k)} \\
e^{-iF(k)} & 0
\end{array} \right);
\end{eqnarray*}
\begin{eqnarray*}
[\textbf{S}_x,\sigma_3]=\frac{2}{q}\left(
\begin{array}{cc}
0 & -1 \\
1 & 0
\end{array} \right)=2\sqrt{T(k)}e^{i(J(k)-kd)}\left(
\begin{array}{cc}
0 & -1 \\
1 & 0
\end{array} \right).
\end{eqnarray*}
This means that in the general case the Shcr\"{o}dinger dynamics crosses the superselection sectors of the Hilbert space associated with this
scattering process.

There are only two particular cases when one of these two scattering matrices commutes with $\sigma_3$. Namely, this takes place when either
reflection coefficient $R(k)$ or the transmission coefficient $T(k)$ is zero. In the first case, which is associated with the resonant
transmission of a particle through the potential barrier, $[\textbf{S}_k,\sigma_3]=0$: in this case the coherent sectors $\Phi_{as}(\Omega_k^+)$
and $\Phi_{as}(\Omega_k^-)$ are invariant with respect to the Shcr\"{o}dinger dynamics. In the second case, which is associated with the full
reflection of a particle off the ideally opaque potential barrier, $[\textbf{S}_x,\sigma_3]=0$: in this case the coherent sectors
$\Phi_{as}(\Omega_x^+)$ and $\Phi_{as}(\Omega_x^-)$ remain invariant in the course of the scattering process.

Let us consider the general case for the unilateral one-dimensional scattering, when there is only one source of particles located, for example,
to the left of the barrier, that is, in the region $\cal{A}$. Now
\begin{eqnarray} \label{14}
\Psi_{in}(x,t)=\int_{-\infty}^\infty A_{L,in}(k)\chi_{k}^{(1)}(x)e^{-iE(k)t/\hbar} dk;\ooo \Psi_{out}(x,t)=\Psi_{L,out}(x,t)+\Psi_{R,out}(x,t),\ppp\\
\Psi_{L,out}=\int_{-\infty}^\infty A_{L,in}(k)\frac{p^*(k)}{q(k)}\chi_{-k}^{(1)}(x)e^{-iE(k)t/\hbar} dk,\ooo \Psi_{R,out}=\int_{-\infty}^\infty
A_{L,in}(k)\frac{1}{q(k)}\chi_{-k}^{(2)}(x)e^{-iE(k)t/\hbar} dk;\nonumber
\end{eqnarray}
$\langle\Psi_{in}|\Psi_{in}\rangle=\overline{T}+\overline{R}=1$ where $\overline{T}=\langle\Psi_{R,out}|\Psi_{R,out}\rangle$,
$\overline{R}=\langle\Psi_{L,out}|\Psi_{L,out}\rangle$.

In this case the in asymptote $\Psi_{in}$ has only the left component that belongs to the coherent sectors $\Phi_{as}(\Omega_x^-)$ and
$\Phi_{as}(\Omega_k^+)$. As regards the out asymptote, it has two components: its left component $\Psi_{L,out}$ belongs to the coherent sectors
$\Phi_{as}(\Omega_x^-)$ and $\Phi_{as}(\Omega_k^-)$, while its right component $\Psi_{R,out}$ belongs to the coherent sectors
$\Phi_{as}(\Omega_x^+)$ and $\Phi_{as}(\Omega_k^+)$. Thus, in line with the SSR, the in asymptote $\Psi_{in}$ is a pure state, while the out
asymptote $\Psi_{out}$, as a coherent superposition of two pure states from different (both in the $x$-space and in the $k$-space) coherent
sectors, is a mixed state. That is, in the course of this process the Shcr\"{o}dinger dynamics converts a pure (initial) state into a mixed
(final) state.

We have to stress that the SSR associated with this scattering process does not forbid superpositions of states from different coherent sectors.
It only forbids to treat them as pure states and to apply Born's interpretation of pure states to such superpositions. Thus, the unilateral
scattering like the bilateral scattering is a "mixture" of two alternative subprocesses, and a complete quantum model of this process must provide
the way of tracing its ("pure") subprocesses at all stages of scattering.

This model must take into account that the in and out asymptotes of each subprocess evolve within a single coherent sector, either in the $x$- or
$k$-space. For example, for the process with the asymptotes (\ref{14}), the in and out asymptotes of the transmission subprocess evolve in a
single coherent sector ($\Phi_{as}(\Omega_k^+)$) in the $k$-space, while those of the reflection subprocess evolve within a single coherent sector
($\Phi_{as}(\Omega_x^-)$) in the $x$-space. At the same time the in and out asymptotes of the transmission subprocess lie in the different
coherent sectors ($\Phi_{as}(\Omega_x^-)$ and $\Phi_{as}(\Omega_x^+)$) in the $x$-space, and those of the reflection subprocess lie in the
different coherent sectors ($\Phi_{as}(\Omega_k^+)$ and $\Phi_{as}(\Omega_k^-)$) in the $k$-space.

\section{Conclusion}

It is shown that in a quantum description of scattering a nonrelativistic particle on a one-dimensional potential barrier the unboundedness of the
position operator plays a crucial role and, as a consequence, the well-known Stone-von Neumann theorem is inapplicable to this process -- the
Shcr\"{o}dinger representation associated with this process is reducible. It is shown that there is a dichotomous-context-induced superselection
rule with the Pauli matrix $\sigma_3$ as a superselection operator. It divides the space of asymptotes, both in the coordinate and momentum
representations, into the direct sum of two coherent sectors. Of importance is the fact that the matrix $\sigma_3$ does not commute with the
scattering matrix, what means that the Shcr\"{o}dinger dynamics crosses the coherent sectors in the course of the unilateral one-dimensional
scattering -- the initial {\it pure} state is converted into a final {\it mixed} state. The quantum mechanical formalism developed for pure states
can be applied only to the subprocesses of this scattering process (transmission and reflection).

\end{document}